\documentclass[doublecol]{epl2}
\usepackage{amsmath}
\usepackage{amssymb}
\usepackage{epsfig}
\usepackage{color}
\usepackage{graphicx,amsmath}

\title{General {properties} of phase diagrams of heavy-fermion metals}
\shorttitle{General properties of phase diagrams}
\author{V. R. Shaginyan \inst{1,2}\thanks {Email:
\email{vrshag@thd.pnpi.spb.ru}} \and A. Z. Msezane \inst{2}\and K.
G. Popov\inst{3} \and G. S. Japaridze \inst{2} \and V. A. Khodel
\inst{4,5}} \shortauthor{V.R. Shaginyan \etal} \institute{\inst{1}
Petersburg Nuclear Physics Institute, Gatchina, 188300,
Russia\\\inst{2} Clark Atlanta University, Atlanta, GA 30314, USA\\
\inst{3} Komi Science Center, Ural Division, RAS, Syktyvkar,
167982, Russia\\ \inst{4} Russian
Research Center Kurchatov Institute, Moscow, 123182, Russia\\
\inst{5} McDonnell Center for the Space Sciences \& Department of
Physics, Washington University, St. Louis, MO 63130, USA}


\abstract{We study the temperature-magnetic field $T-B$ phase
diagrams of {{heavy fermion (HF)}} metals, and show that at
sufficiently high temperatures outside the ordered phase the
crossover temperature $T^*(B)$, regarded as the energy scale,
follows a linear $B$-dependence, crossing the origin of the $T-B$
phase diagram. This behavior of $T^*(B)$ constitutes the general
property, and  is formed by the presence of fermion condensation
quantum phase transition hidden within the ordered phase. Our
result is in good agreement with the experimental $T-B$ phase
diagram of the HF metals $\rm YbRh_2Si_2$, $\rm
Yb(Rh_{0.93}Co_{0.07})_2Si_2$, and $\rm
Yb(Rh_{0.94}Ir_{0.06})_2Si_2$. To support our observations, we
analyze the isothermal magnetization $M$, and demonstrate  that
$dM/dT$ exhibits a universal temperature behavior over magnetic
field scaling. The obtained results are in good agreement with the
corresponding data collected on $\rm YbRh_2Si_2$ as a function of
magnetic field at different temperatures under hydrostatic
pressure.}

\begin{document}
\maketitle

\section{Introduction}
\label{intro} Quantum criticality occupies the low temperature area
of the $T-B$ phase diagram of the heavy-fermion (HF) metals. In
this area HF metals exhibit a specific behavior known as non-Fermi
liquid (NFL) behavior, which is induced by a phase transition at
zero temperature, driven by pressure, doping or magnetic field.
While approaching the quantum critical point (QCP), related to the
phase transition, the effective mass $M^*$ grows continuously,
causing strong deviations from Landau's Fermi liquid (LFL) theory.
In the $T-B$ phase diagram, such NFL states are separated from
those of LFL by the crossover temperature $T^*(B)$, given by the
behavior of $M^*$ and regarded as the energy scale, which vanishes
at the QCP \cite{shaginyan:2010}. The QCP can be related to
different quantum phase transitions such as antiferromagnetic (AF)
phase transition, e.g. $\rm YbRh_2Si_2$ \cite{brando:2013},
ferromagnetic (FM) one, e.g. $\rm{CePd_{1-x}Rh_x}$ \cite{sereni},
superconducting (SC), e.g. $\rm CeCoIn_5$ \cite{geg2013}, or even
the QCP can be located at the origin of the $T-B$ phase diagram
with the paramagnetic (PM) ground state, e.g. $\rm {CeRu_2Si_2}$
\cite{takahashi:2003}. Nonetheless, such a diverse transitions do
not violate the universal scaling behavior of the thermodynamic
properties, and they are indicative of the diversity generated by a
single phase transition represented by the fermion condensation
quantum phase transition (FCQPT) \cite{shaginyan:2010}. It is
reasonable to expect that fingerprints of that phase transition
emerge at sufficiently high temperatures $T^*_d>T_{NL}$, where
$T_{NL}$ is the critical temperature of the AF phase transition
which is taken as an example.

Recent study poses a challenging problem for the condensed matter
physics, revealing that the $T-B$ phase diagrams of HF metals $\rm
CeCoIn_5$ \cite{geg2013}, $\rm YbRh_2Si_2$, $\rm
Yb(Rh_{0.93}Co_{0.07})_2Si_2$, and $\rm
Yb(Rh_{0.94}Ir_{0.06})_2Si_2$, and the evolution of the $T-B$
diagram of $\rm YbRh_2Si_2$ under the application of hydrostatic
pressure $P$ indicates a hidden QCP at zero field, while $T^*(B)$
follows a linear $B$-dependence at sufficiently high fields or
temperatures outside the ordered phase \cite{brando:2013,geg2013}.
It has been observed that at sufficiently high temperatures $T$
outside the AF phase the crossover temperature $T^*(B)$ follows
almost a linear $B$-dependence, crossing the origin of the $T-B$
phase diagram \cite{brando:2013}. Phase diagrams for $\rm
Yb(Rh_{0.93}Co_{0.07})_2Si_2$ and $\rm
Yb(Rh_{0.94}Ir_{0.06})_2Si_2$ can be obtained from the phase
diagram for $\rm YbRh_2Si_2$ by applying positive/negative
pressure. The negative chemical pressure is caused by $\rm Ir$
substitution and the positive by $\rm Co$ substitution. This can be
confirmed by performing hydrostatic pressure experiments on clean
undoped $\rm YbRh_2Si_2$ and comparing the results with ambient
pressure ones on $\rm Yb(Rh_{1-x}Co_x)_2Si_2$, demonstrating that
the $\rm Co$ and $\rm Ir$-induced disorder can not be the reason
for the modifications of the corresponding $T-B$ phase diagram
\cite{brando:2013,tokiwa:2009:A}. Thus, the phase diagrams obtained
in measurements on $\rm Yb(Rh_{0.93}Co_{0.07})_2Si_2$ and $\rm
Yb(Rh_{0.94}Ir_{0.06})_2Si_2$ can be viewed as a general
experimental $T-B$ phase diagram of $\rm YbRh_2Si_2$ that allows
one to reveal QCP hidden in the AF phase.

In this letter, to resolve the challenging problem mentioned above,
we study the $T-B$ phase diagrams of HF metals, and show that at
sufficiently high temperatures outside the ordered phase the
crossover temperature $T^*(B)$ follows a linear $B$-dependence, and
crosses the origin of the $T-B$ phase diagram. Upon analyzing the
experimental global $T-B$ phase diagram of $\rm YbRh_2Si_2$, we
show that FCQPT represents QCP hidden in the AF phase. Our analysis
agrees well with the experimental $T-B$ phase diagrams of the HF
metals $\rm YbRh_2Si_2$, $\rm Yb(Rh_{0.93}Co_{0.07})_2Si_2$, and
$\rm Yb(Rh_{0.94}Ir_{0.06})_2Si_2$. We calculate the isothermal
magnetization $M$, and demonstrate  that $dM/dT$ exhibits a
universal temperature over magnetic field scaling. Our results are
in good agreement with data collected on $\rm YbRh_2Si_2$, and
support our conclusion on the nature of the hidden QCP.

\section{Energy scales and phase diagrams of HF metals}
\label{HCEL9}

\begin{figure}
\begin{center}
\vspace*{-0.2cm}
\includegraphics [width=0.47\textwidth]{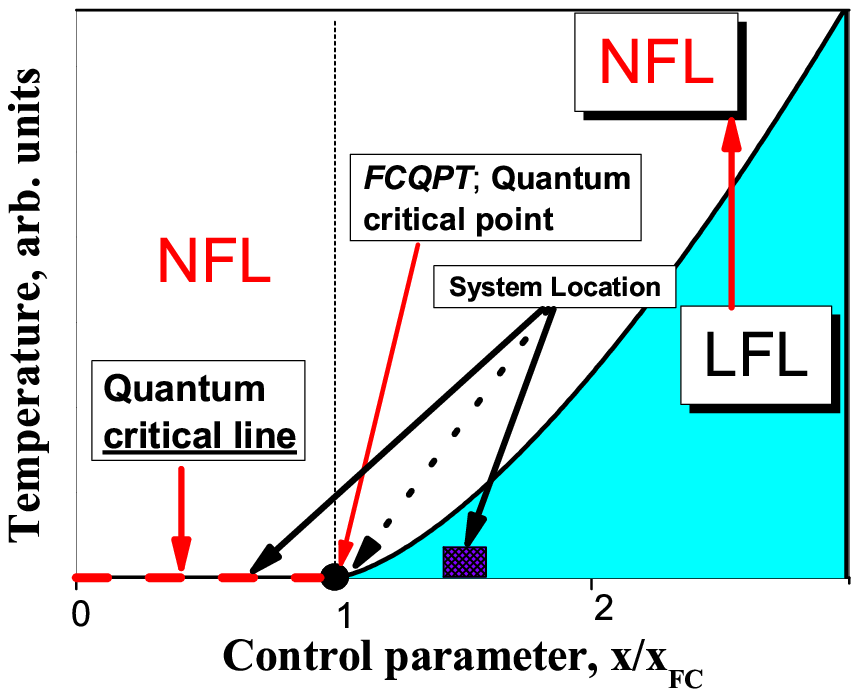}
\end{center}
\vspace*{-0.3cm} \caption{(Color online) Schematic $T-x$ phase
diagram of system with FC. The number density $x$ is taken as the
control parameter and depicted as $x/x_{FC}$. At $x/x_{FC}>1$ and
at sufficiently low temperatures, the system is in the LFL state as
shown by the shadowed area. This location of the system is depicted
by both the solid square and the arrow. The vertical arrow
illustrates the system moving in the LFL-NFL direction along $T$ at
fixed control parameter.  At $x/x_{FC}<1$ the system is shifted
beyond the QCP, and is at the quantum critical line depicted by the
dashed line and shown by the vertical arrow. At any finite low
temperatures $T>0$ the system possesses finite entropy $S_0$ and
exhibits the NFL behavior.}\label{fig03bel}
\end{figure}

\begin{figure}
\begin{center}
\includegraphics [width=0.47\textwidth]{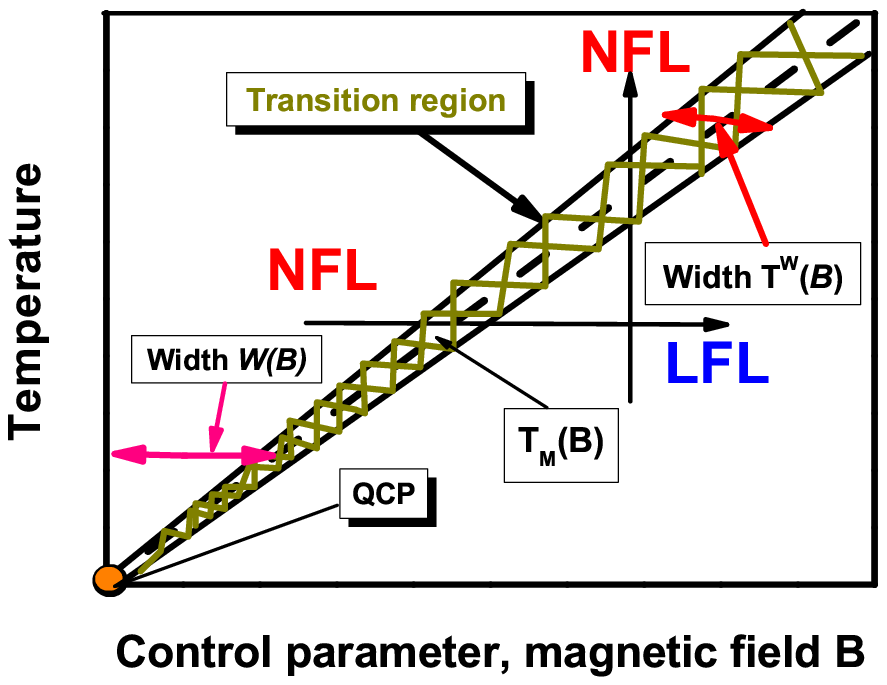}
\end{center}
\caption{(Color online) Schematic $T-B$ phase diagram of HF liquid
with magnetic field as the control parameter. The vertical and
horizontal arrows show LFL-NFL and NFL-LFL transitions at fixed $B$
and $T$, respectively. At $B=0$ the system is in its NFL state,
having a flat band, and demonstrates NFL behavior down to $T\to0$.
The hatched area separates the NFL phase and the weakly polarized
LFL phase and represents the transition region. The dashed line in
the hatched area represents the function $T_M(B)\simeq T^*_{FC}$
given by $T_M\simeq a_1\mu_BB$. The functions $W(B)\propto T$ and
$T^W(B)\propto T$ shown by two-headed arrows define the width of
the NFL state and the transition area, respectively. The QCP
located at the origin and indicated by the arrow denotes the
critical point at which the effective mass $M^*$ diverges and both
$W(B)$ and $T^W(B)$ tend to zero.}\label{fig1bel}
\end{figure}

At $T=0$, a quantum phase transition is driven by a nonthermal
control parameter such as number density $x$, magnetic field $B$ or
pressure $P$. At the QCP, situated at $x=x_{FC}$ and related to
FCQPT, the effective mass $M^*$ diverges. We note that there are
different types of instabilities of normal Fermi liquids related to
perturbations of the initial quasiparticle spectrum
$\varepsilon(p)$ and occupation numbers $n(p)$, associated with the
emergence of a multi-connected Fermi surface, see e.g.
\cite{khodel:2008,shaginyan:2010}. Depending on the parameters and
analytical properties of the Landau interaction, such instabilities
lead to several possible types of restructuring of the initial
Fermi liquid ground state. In fact, at elevated temperatures the
systems located at these transition points, exhibit the behavior
typical to those located at FCQPT \cite{shaginyan:2010}. Therefore,
we do not consider the specific properties of systems located at
different topological transitions, but rather focus on the behavior
of the system located near FCQPT. Beyond the FCQPT, the system
shapes FC that leads to the formation of a topologically protected
flat band \cite{volovik:1991,volovik:1994}. {We note that
microscopic analysis confirms that the formation is robust when
tuning interaction, temperature, and chemical potential
\cite{yudin:2014}, and these results are in accordance with
\cite{volovik:1991,volovik:1994}.

For the reader convenience, and to present a coherent analysis of
the $T^*(B)$ behavior based on the recent experimental facts
\cite{brando:2013,tokiwa:2009:A}, we start with consideration of
the auxiliary phase diagrams reported in figs. \ref{fig03bel},
\ref{fig1bel}, \ref{fig04bel}, and \ref{f02bel} which were partly
considered in Refs. \cite{shaginyan:2010,shaginyan:2012}}. The
schematic $T-x$ phase diagram of the system driven to the FC state
by varying the number density $x$ is presented in fig.
\ref{fig03bel}. Upon approaching the critical density $x_{\rm FC}$
the system remains in the LFL region at sufficiently low
temperatures, as shown by the shadowed area. The temperature range
of this area shrinks as the system approaches QCP, and $M^*(x\to
x_{FC})$ diverges. At this QCP shown by the arrow in fig.
\ref{fig03bel}, the system demonstrates the NFL behavior down to
the lowest temperatures. Beyond the critical point the behavior
remains NFL  even at $T\to0$. It is determined by the
temperature-independent residual entropy $S_0$
\cite{shaginyan:2010,khodel:2005}. In that case at $T\to 0$, the
system approaches a quantum critical line (QCL) (shown by the
vertical arrow and the dashed line in fig. \ref{fig03bel}) rather
than a QCP. Upon reaching the QCL from above at $T\to 0$ the system
undergoes the first order quantum phase transition, making the
residual entropy $S_0$ vanish. As seen from fig. \ref{fig03bel}, at
rising temperatures the system located before QCP does not undergo
a phase transition, and transits from the NFL to the LFL regime. At
finite temperatures there is no boundary (or phase transition)
between the states of the system located before or behind QCP,
shown by the arrows. Therefore, at elevated temperatures the
properties of systems with $x/x_{\rm FC}<1$ or with $x/x_{FC}>1$
become indistinguishable.

As seen from fig. \ref{fig03bel}, the location of the system is
controlled by the number density $x$. At $x/x_{FC}>1$, the system
is located before FCQPT, and demonstrates the LFL behavior at low
temperatures. We speculate that such a state can be induced by
applying positive pressure, including positive chemical pressure,
as occurs for the case of $\rm Yb(Rh_{0.93}Co_{0.07})_2Si_2$. HF
metals, such as  $\rm {CeRu_2Si_2}$ are placed at FCQPT, as shown
in fig. \ref{fig03bel} by the dash arrow. Such metals exhibit PM
ground state and NFL behavior to the lowest temperatures. On the
other hand, by diminishing $x$, $x/x_{FC}<1$, the system is shifted
beyond FCQPT, and is at the quantum critical line depicted by the
dashed line. In that case the system demonstrates the NFL behavior
at any finite temperatures. We assume that such a state can be
induced by the application of negative pressure, including negative
chemical pressure, as is the case for $\rm
Yb(Rh_{0.94}Ir_{0.06})_2Si_2$. At low temperatures and above the
critical line, the system has the finite entropy $S_0$ and its NFL
state is strongly degenerate. The degeneracy stimulates the
emergence of different phase transitions, lifting degeneracy and
removing the entropy $S_0$. The NFL state can be captured by other
states such as SC (like SC state in $\rm CeCoIn_5$), or AF (like
the AF state in $\rm YbRh_2Si_2$)
\cite{shaginyan:2010,khodel:2005}. The diversity of phase
transitions at low temperatures is one of the most spectacular
features of the physics of many HF metals and strongly correlated
compounds. Within the scenario of ordinary quantum phase
transitions, it is hard to understand why these transitions are so
different from each other and their critical temperatures are so
extremely small. However, such diversity is endemic to systems with
a FC, since the FC state should be altered as $T\to 0$ so that the
excess entropy $S_0$ is shed before zero temperature is reached. At
finite temperatures this takes place by means of some phase
transitions which can compete, shedding the excess entropy
\cite{shaginyan:2010,khodel:2008}.

The schematic $T-B$ phase diagram of a HF liquid is presented in
fig.~\ref{fig1bel}, with the magnetic field $B$ serving as control
parameter. The NFL regime reigns at elevated temperatures and fixed
magnetic field.  With $B$ increasing, the system is driven from the
NFL region to the LFL domain. The magnetic-field-tuned QCP is
indicated by the arrow and located at the origin of the phase
diagram, since application of any magnetic field destroys the flat
band and shifts the system into the LFL state
\cite{shaginyan:2010,takahashi:2003,gegenwart:2002}. Near the
magnetic-field-tuned QCP, a deeper insight into the behavior of
$M^*(B,T)$ can be achieved using some "internal" (or natural)
scales. Namely, $M^*(B,T)$ reaches its maximum value {$M^*_{max}$
at some temperature $T_M\simeq a_1\mu_BB$, where $a_1$ is a
dimensionless factor and $\mu_B$ is the Bohr magneton, see e.g.
\cite{shaginyan:2010}. It is convenient to introduce the internal
scales { {$M^*_{max}$}} and $T_{M}$ to measure the effective mass
and temperature. Thus, we divide the effective mass $M^*$ and the
temperature $T$ by the values, { {$M^*_{max}$}} and $T_{M}$,
respectively. This generates the normalized effective mass {
{$M^*_N=M^*/M^*_{max}$}} and the normalized temperature
$T_N=T/T_{M}$. $M^*_N(T_N)$ can be well approximated by a simple
universal interpolating function \cite{shaginyan:2010}. The
interpolation is valid between the LFL and NFL regimes and
represents the universal scaling behavior of $M^*_N$
\begin{equation}M^*_N(y)\approx c_0\frac{1+c_1y^2}{1+c_2y^{8/3}}.
\label{UN2}
\end{equation}
Here, $y=T_N=T/T_{M}$, $c_0=(1+c_2)/(1+c_1)$, $c_1$, $c_2$ are
fitting parameters. Thus, in the presence of magnetic field eq.
\eqref{UN2} describes the scaling behavior of the effective mass as
a function of $T$ versus $B$ - the curves $M^*_{N}$ at different
magnetic fields $B$ merge into a single one in terms of the
normalized variable $y=T/T_M$. Since the variables $T$ and $B$
enter symmetrically, eq. \eqref{UN2} describes the scaling behavior
of $M^*_{N}(B,T)$ as a function of $B$ versus $T$. In fig.
\ref{fig1bel}, the hatched area denoting the transition region
separates the NFL state from the weakly polarized LFL state and
contains the dashed line tracing the transition region,
$T_M(B)\simeq T^*_{FC}$. Referring to eq.~\eqref{UN2}, this line is
defined by the function $T^*_{FC}\propto \mu_BB$, and the width
$W(B)$ of the NFL state is seen to be proportional to $T$.
Similarly, it can be shown that the vertical width $T^W(B)$ of the
transition region is also proportional to $T$.

\begin{figure}
\begin{center}
\includegraphics [width=0.47\textwidth]{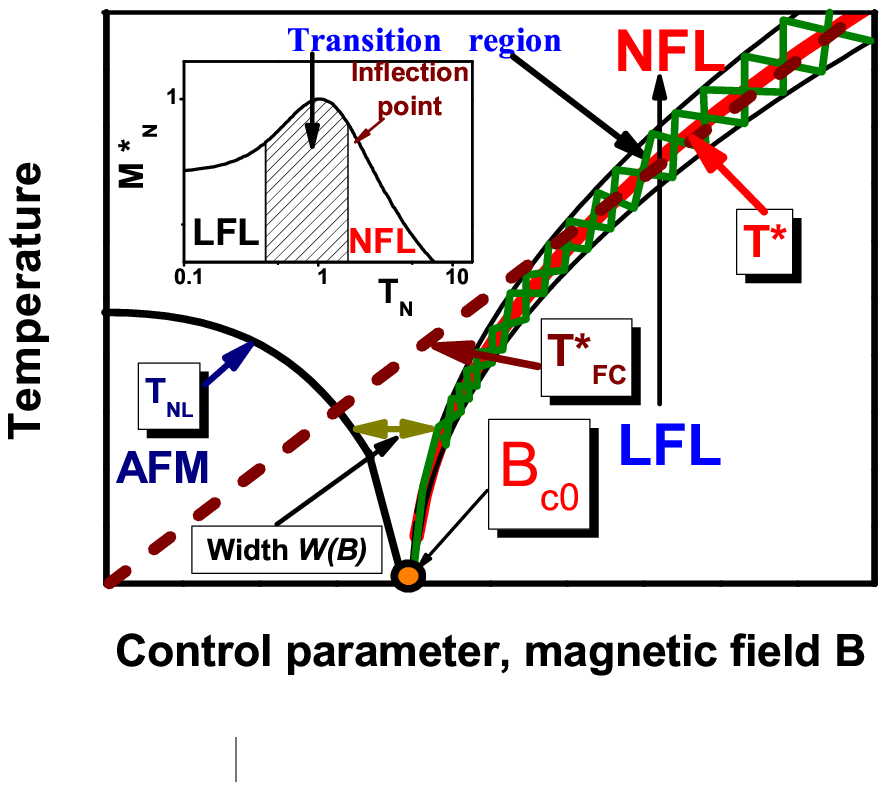}
\vspace*{-1.5cm}
\end{center}
\caption{(Color online) Schematic $T-B$ phase diagram of a HF metal
with magnetic field as control parameter. The AF phase boundary
line is shown by the arrow and plotted by the solid curve,
representing the N\'eel temperature $T_{NL}$. The hatched area
separates the NFL phase and the weakly polarized LFL one and
represents the transition region. The solid curve inside the
hatched area represents the transition temperature $T^*$. The
dashed line $T^*_{FC}(B)\propto B\mu_B$ shows the transition
temperature provided that the AF state were absent. The function
$W(B)\propto T$ shown by the two-headed {{arrow}} defines the total
width of both the NFL state and the transition area. The inset
shows a schematic plot of the normalized effective mass versus the
normalized temperature. The transition region, where $M^*_N$
reaches its maximum at $T_N=T/T_M=1$, is shown as the hatched area
in both the main panel and the inset. Arrows indicate the
transition region and the inflection point $T_{inf}$ in the $M^*_N$
plot.}\label{fig04bel}
\end{figure}
\begin{figure}
\begin{center}
\vspace*{-0.2cm}
\includegraphics [width=0.47\textwidth]{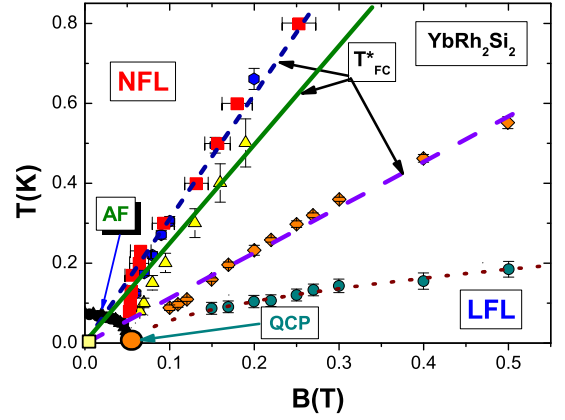}
\end{center}
\vspace*{-0.3cm} \caption{(Color online) $T-B$ phase diagram for
$\rm YbRh_2Si_2$. Solid circles represent the boundary between AF
and NFL states. The solid circles along the dotted curve denote the
boundary of the NFL and LFL regions
\cite{gegenwart:2007,gegenwart:2002}. The solid square at the
origin represents FCQPT. The dotted curve represents the function
$\sqrt{B-B_{c0}}$ \cite{shaginyan:2010}. The solid line represents
$T^*_{FC}\propto B\mu_B$, and shows the linear fit to $T^*(B)$.
{The emphasis is placed on the relatively high temperatures,
therefore, the lines $T^*_{FC}$ deviate from the experimental
points at decreasing temperatures.} Diamonds marking $T^*$ along
the dash line signify the maxima $T_M$ of $C/T$
\cite{oeschler:2008}. The transition temperature $T^*$, determined
from magnetostriction (solid squares and the short dash line),
longitudinal magnetoresistivity (triangles and the solid line), and
susceptibility (solid circles and the short dash line)
\cite{gegenwart:2007} is also reported.}\label{f02bel}
\end{figure}
We now construct the $T-B$ schematic phase diagram of a HF metal
like $\rm YbRh_2Si_2$ shown in fig. \ref{fig04bel}. The inset in
fig. \ref{fig04bel} demonstrates the scaling of the normalized
effective mass $M^*_N$ versus the normalized temperature $T_N$. The
LFL phase prevails at $T\ll T_M$, followed by the $T^{-2/3}$ regime
at $T \gtrsim T_M$. The latter phase is designated as NFL due to
the strong temperature dependence of the effective mass in it. The
region $T\simeq T_M$ encompasses the transition between the LFL
regime with almost constant effective mass and the NFL behavior.
When constructing the phase diagram in fig.~\ref{fig04bel}, we
assumed that  the AF order prevails, destroying the $S_0$ term at
low temperatures. At $B=B_{c0}$, the HF liquid acquires a flat
band, where $B_{c0}$ is a critical magnetic field, such that at
$T\to 0$ the application of magnetic field $B\gtrsim B_{c0}$
destroys the AF state restoring the PM one with LFL behavior. In
some cases $B_{c0}=0$ as in the HF metal $\rm CeRu_2Si_2$ (see e.g.
\cite{takahashi:2003}), while in $\rm YbRh_2Si_2$, $B_{c0}\simeq
0.06$ T \cite{gegenwart:2002}. Obviously, $B_{c0}$ is defined by
the specific system properties; therefore we consider it as a
parameter. At elevated temperature and fixed magnetic field the NFL
regime is dominant. With $B$ increasing, the system is driven from
the NFL to the LFL domain. The magnetic-field-tuned QCP is
indicated by the arrow and is located at $B=B_{c0}$. The hatched
area denotes the transition region, and separates the NFL state
from the weakly polarized LFL one. This area contains both the
dashed line tracing $T^*_{FC}(B)$ and the solid curve $T^*(B)$.
Referring to eq. \eqref{UN2}, the latter is defined by the function
$T^*\propto\mu_BB$ and merges with $T^*_{FC}(B)$ at relatively high
temperatures, and $T^*\propto\mu_B(B-B_{c0})$ at lower $T\sim
T_{NL}$, with $T_{NL}(B)$ being the N\'eel temperature. As seen
from eq. \eqref{UN2}, both the width $W(B)$ of the NFL state and
the width of the transition region are proportional {{to}} $T$. The
AF phase boundary line is shown by the arrow and depicted by the
solid curve. As mentioned above, the dashed line
$T^*_{FC}(B)\propto B\mu_B$ represents the transition temperature
provided that the AF state were absent. In that case the FC state
is destroyed by any weak magnetic field $B\to0$ at $T\to0$ and the
dashed line $T^*_{FC}$ crosses the origin, as is displayed in fig.
\ref{fig04bel}. At $T\gtrsim T_{NL}(B=0)$ the transition
temperature $T^*_{FC}(B)$ coincides with $T^*(B)$ shown by the
solid curve, since the properties of the system are given by its
local free energy, describing the PM state of the system. One might
say that the system "does not remember" the AF state, emerging at
lower temperatures. This observation is in good agreement with the
data collected on the HF metal $\rm YbRh_2Si_2$.

\section{The HF metal $\rm YbRh_2Si_2$ under chemical and hydrostatic pressure}

The above findings are summarized in the phase diagram of fig.
\ref{f02bel}. At relatively high temperatures $T\gtrsim
T_{NF}(B=0)$ the transition temperature $T^*$, obtained in
measurements on $\rm YbRh_2Si_2$
\cite{gegenwart:2007,gegenwart:2002,oeschler:2008}, is well
approximated by the straight lines representing $T^*_{FC}$. It is
seen that the straight lines deviate from the experimental points
at relatively low temperatures. Also from fig. \ref{f02bel}, the
slope of the dash line (representing the maxima of the specific
heat $C/T$) is different from that of the short dash line
(representing maxima of the susceptibility $\chi(T)$). Such
behavior is determined by the fact that the maxima of $C/T$ and
$\chi(T)$ are defined from the two different relations, determining
the inflection points of the entropy:
\begin{equation}
 \frac{\partial^2S}{\partial T^2} = 0, \quad
\frac{\partial^2S}{\partial B\,^2} = 0,\label{inf_TB}
\end{equation}
respectively. The theory of FC shows that the inflection points
{{exist}} under the condition that the system is located near FCQPT
\cite{shaginyan:2010}.

\begin{figure}
\begin{center}
\vspace*{-0.2cm}
\includegraphics [width=0.47\textwidth]{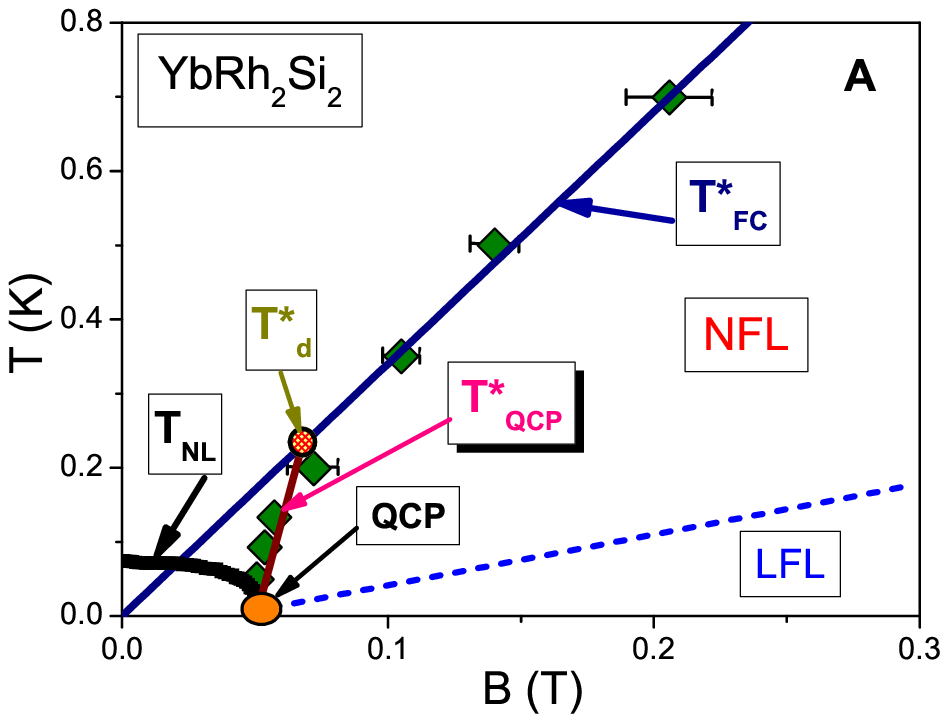}
\vspace*{-0.2cm}
\includegraphics [width=0.47\textwidth]{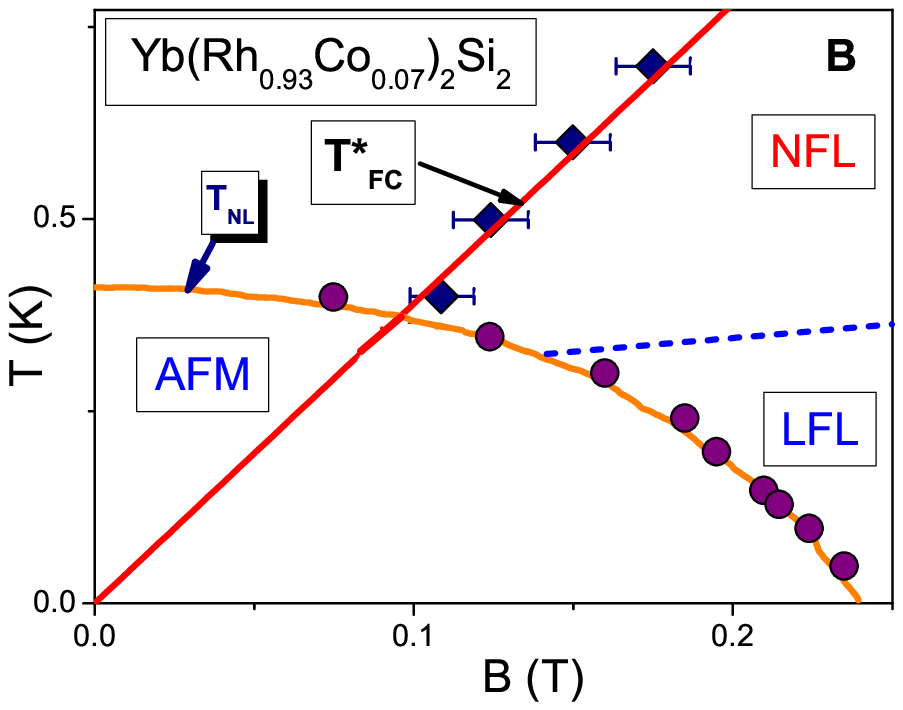}
\vspace*{-0.2cm}
\includegraphics [width=0.47\textwidth]{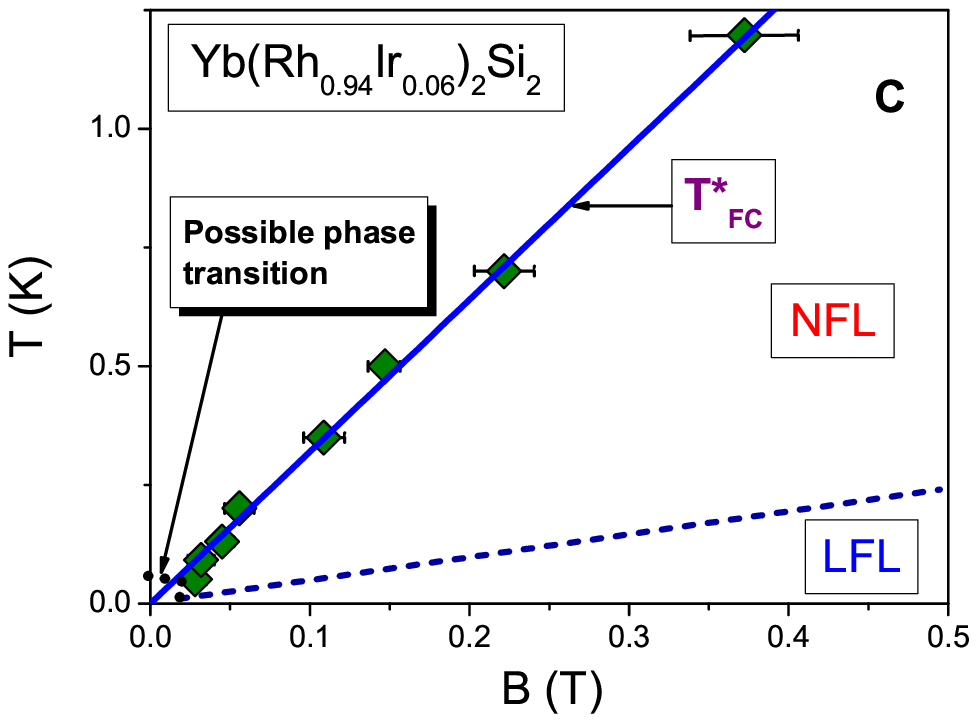}
\end{center}
\vspace*{-0.3cm} \caption{(Color online) $T-B$ phase diagrams for
the three HF metals: $\rm YbRh_2Si_2$ (A); $\rm
Yb(Rh_{0.93}Co_{0.07})_2Si_2$ (B); $\rm
Yb(Rh_{0.94}Ir_{0.06})_2Si_2$ (C). In panels A and B, the AF phase
boundaries \cite{brando:2013} are shown by the solid lines. The
diamonds correspond to the measurements of $T^*(B)$ extracted from
the analysis of the $\hat{\bf{M}}(B)$ function \cite{brando:2013}.
$T_d$ is the temperature at which the line $T^*_{FC}$ starts to
deviate from the experimental data. In panel C, the phase boundary
of possible phase transition is shown by dotted curve. The solid
straight lines depict the transition temperature $T^*_{FC}(B)$. The
short dash lines represent schematically the boundary between NFL
and LFL regions.}\label{ff4bel}
\end{figure}
Panels A, B, C of fig. \ref{ff4bel} are focused on the behavior of
the transition temperature $T^*(B)$ extracted from measurements of
kinks in $\hat{\textbf{M}}(B)=\textbf{M}+B(d\textbf{M}/dB)$
\cite{brando:2013}, where $\textbf{M}$ is the magnetization.
Positions of the kinks are represented by diamonds in fig.
\ref{ff4bel}. It is seen from the panel A of fig. \ref{ff4bel},
that at $T\gtrsim T_{NL}(B=0)$, the transition temperature $T^*$ of
$\rm YbRh_2Si_2$ is well approximated by the line $T^*_{FC}$.
$T_d\sim B_{c0}$ is the temperature at which the line $T^*_{FC}$
starts to deviate from the experimental data. Indeed, it is seen
from fig. \ref{fig1bel} that the width $W\sim T^W\sim T\sim B$.
Approaching QCP at $B\sim B_{c0}$, as seen from fig.
\ref{fig04bel}, the width narrows, $W\sim(B-B_{c0})$, and
$T^*_{FC}$ deviates from $T^*$. As mentioned above, upon using
nonthermal tuning parameters like the number density $x$, the NFL
regime may disappear and the LFL is restored. In our simple model,
applying positive pressure $P$ leads to the increase of the density
$x$ which in turn leads to transition form FCQPT to the LFL state.
Figure \ref{fig03bel} shows that the above actions move the
electronic system of $\rm YbRh_2Si_2$ into the shadowed area
characterized by the LFL behavior at low temperatures. The new
location of the system, represented by $\rm
Yb(Rh_{0.93}Co_{0.07})_2Si_2$, is shown by the arrow pointing at
the solid square. We note that the positive chemical pressure in
the considered case is induced by $\rm Co$ substitution
\cite{tokiwa:2009:A,brando:2013}. As a result, the application of
magnetic field $B\simeq B_{c0}$ does not drive the system to its
FCQPT with the divergent effective mass because the QCP has already
been destroyed by the positive pressure, as shown in panel B of
fig. \ref{ff4bel}. Here $B_{c0}$ is the critical magnetic field
that eliminates the corresponding AF order. At $B<B_{c0}$ and
diminishing temperatures, the system enters the AF state with the
LFL behavior. As a result, the effective mass given by eq.
\eqref{UN2} becomes $T$-independent and the width $W(B)$ remains
constant inside the AF phase. At $B>B_{c0}$ and increasing
temperature, the system, moving along the vertical arrow, transits
from the LFL regime to the NFL one. At relatively high temperatures
both $\rm YbRh_2Si_2$ and $\rm Yb(Rh_{0.93}Co_{0.07})_2Si_2$ are in
their PM states. As a result, $T^*$ is well approximated by the
straight line $T^*_{FC}$. This behavior coincides with the
experimental data \cite{tokiwa:2009:A,brando:2013} reported on
panels A, B of fig. \ref{ff4bel}. The system located above QCL
exhibits the NFL behavior down to the lowest temperatures unless it
is captured by a phase transition. The behavior exhibited by the
system located above QCL agrees with the experimental observations
of the QCP evolution in $\rm YbRh_2Si_2$ under the application of
negative chemical pressure induced by Ir substitution
\cite{tokiwa:2009:A,brando:2013}. To explain the latter behavior,
we propose a simple model that the application of negative pressure
reduces $x$ so that the electronic system of $\rm YbRh_2Si_2$ moves
from QCP to a new position over QCL shown by the dash arrow in fig.
\ref{fig03bel}. Thus, the electronic system of $\rm
Yb(Rh_{0.94}Ir_{0.06})_2Si_2$ is located at QCL and possesses a
flat band, while the entropy includes $S_0$. We predict that at
temperature decreasing, the electronic system of $\rm
Yb(Rh_{0.94}Ir_{0.06})_2Si_2$ is captured by a phase transition,
since the NFL state above QCL is strongly degenerate and the term
$S_0$ has to vanish. As temperature decreases, this degeneracy is
lifted by some phase transition which can be possibly detected
through the LFL state accompanying it. The tentative boundary line
of that transition is shown by the short dashed line in fig.
\ref{ff4bel}, panel C. It is also seen from panel C of fig.
\ref{ff4bel}, that at elevated temperatures $T^*$ is well
approximated by the function $T^*_{FC}$. Thus, at relatively high
temperatures the curve $T^*_{FC}(B)$, shown in panels A, B, C of
fig. \ref{ff4bel} by the solid lines, coincides with $T^*(B)$
depicted by the diamonds. The preceding discussion demonstrates
that the local properties of the system are given by its local free
energy. This local free energy defines the NFL behavior which is
formed by the presence of FQCPT, as it is reported in  fig.
\ref{fig03bel}.

To confirm the above consideration of the phase diagrams, we
describe both low temperature magnetization measurements carried
out under pressure $P$ and the $T-B$ phase diagram of $\rm
YbRh_2Si_2$ near its magnetic-field-tuned QCP
\cite{tokiwa:2009,tokiwa:2009:A}. To carry out a quantitative
analysis of the scaling behavior of $-\Delta M^*(B,T)/\Delta T$, we
calculate the entropy $S(B,T)$ and employ the well-known
thermodynamic equality $dM/dT=dS/dB\simeq\Delta M/\Delta T$
\cite{shaginyan:2010,tokiwa:2009,tokiwa:2009:A}. Figure
\ref{figMTPD}, panel a, reports the normalized $(dS/dB)_N$ as a
function of the normalized magnetic field. The function $(dS/dB)_N$
is obtained by normalizing $(dS/dB)$ by its maximum taking place at
$B_M$, and the field $B$ is scaled by $B_M$. It is seen from fig.
\ref{figMTPD}, panel a, that our calculations are in good agreement
with the experimental data and the fitting functions $-(\Delta
M/\Delta T)_N$ show the scaling behavior over three decades in the
normalized magnetic field.
\begin{figure}
\begin{center}
\vspace*{-0.2cm}
\includegraphics [width=0.49\textwidth]{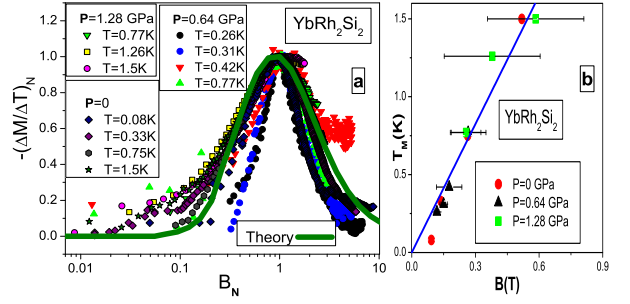}
\end{center}
\vspace*{-0.3cm} \caption{(Color online) Panel a: The normalized
magnetization difference divided by temperature increment $-(\Delta
M/\Delta T)_N$ vs normalized magnetic field at fixed temperature
and pressure (listed in the legend in the upper left corner) is
extracted from the data collected  on $\rm YbRh_2Si_2$
\cite{tokiwa:2009,tokiwa:2009:A}. Panel b: $T_M$ versus $B$ is
obtained under the application of hydrostatic pressure depicted in
the legend and extracted from the measurements
\cite{tokiwa:2009,tokiwa:2009:A}. The solid line shows the linear
fit to $T^*(B)$, and represents $T^*_{FC}$.}\label{figMTPD}
\end{figure}
fig. \ref{figMTPD}, panel b, presents the temperature $T_M$, at
which the maximum $-(\Delta M/\Delta T)$ takes place, as a function
of magnetic field $B$. At $T_M\simeq T_d$ the line $T^*_{FC}$
starts to deviate from the experimental data, where $T_d$ is shown
in fig. \ref{ff4bel}, panel A.

\section{Summary and acknowledgements}

We have carried out a comprehensive theoretical study of the $T-B$
phase diagrams of HF metals such as $\rm YbRh_2Si_2$ and considered
the evolution of these diagrams under the application of
negative/positive pressure. We have observed that at sufficiently
high temperatures outside the AF phase the transition temperature
$T^*(B)$ follows a linear $B$-dependence, and collapses on the
universal line $T^*_{FC}(B)$  through the origin. This behavior is
in agreement with experimental facts, signifies the general
properties of the phase diagram, and is induced by the presence of
FCQPT hidden within the AF phase.

This work was partly supported by the RFBR \# 14-02-00044, U.S.
DOE, Division of Chemical Sciences, Office of Energy Research and
AFOSR.


\begin{thebibliography}{10}

\bibitem{shaginyan:2010}
\Name{Shaginyan V. R., Amusia M. Ya., Msezane A. Z. \and Popov K.
G.}\REVIEW{Phys. Rep.}{492}{2010}{31}.

\bibitem{brando:2013}
\Name{Brando M. \etal} \REVIEW{Phys. Status Solidi B} {459}{2013}
{285}.

\bibitem{sereni} \Name{ Sereni J. G. \etal} \REVIEW{Phys. Rev. B}
{75}{2007}{024432}.

\bibitem{geg2013} \Name{ Tokiwa Y., Bauer E. D. and P. Gegenwart} \REVIEW{Phys.
Rev. Lett.} {111}{2013}{107003}.

\bibitem{takahashi:2003} \Name{Takahashi D. \etal} \REVIEW{Phys. Rev. B}{67}{2003}{180407(R)}.

\bibitem{tokiwa:2009:A}
\Name{Tokiwa Y. \etal} \REVIEW{J. Phys. Soc. Jpn.}
  {78}{2009}{123708}.

\bibitem{khodel:2008}
\Name{Khodel V.A., Clark J.W. \and Zverev M.V.} \REVIEW{Phys. Rev.
B}{78}{2008}{075120}.

\bibitem{volovik:1991}
\Name{Volovik G. E.}  \REVIEW{JETP Lett.} {53}{1991}{222}.

\bibitem{volovik:1994} {\Name{Volovik G. E.} \REVIEW{JETP Lett.} {59}{1994}{830}}.

\bibitem{yudin:2014} {\Name{Yudin D. \etal} \REVIEW{Phys. Rev.
Lett.} {112}{2014}{070403}}.

\bibitem{shaginyan:2012} {\Name{Shaginyan V. R. \etal}
\REVIEW{JETP Lett.} {96}{2012}{397}.}

\bibitem{khodel:2005}
\Name{Khodel V. A., Zverev M.V. \and Yakovenko V. M.} \REVIEW{Phys.
Rev. Lett.} {95}{2005}{236402}.

\bibitem{gegenwart:2002}
\Name{Gegenwart P. \etal} \REVIEW{Phys. Rev. Lett.} {89}{2002}
{056402}.

\bibitem{gegenwart:2007}
\Name{Gegenwart P. \etal} \REVIEW{Science}{315}{2007}{969}.

\bibitem{oeschler:2008}
\Name{Oeschler N. \etal} \REVIEW{Physica B}{403}{2008}{1254}.

\bibitem{tokiwa:2009}
\Name{Tokiwa Y. \etal} \REVIEW{Phys. Rev. Lett.}
{102}{2009}{066401}.

\end{thebibliography}
\end{document}